\documentclass[twocolumn,showpacs,preprintnumbers,amsmath,amssymb]{revtex4-1}

\usepackage[utf8]{inputenc}
\usepackage{amssymb}
\usepackage{graphicx}
\usepackage{dcolumn}
\usepackage{bm}
\usepackage{subfigure}
\usepackage{color}
\usepackage{hyperref}

\hypersetup{
  urlcolor= blue,        
  linkcolor= blue,        
  colorlinks=true,       
  linkcolor=blue, 
  citecolor=blue,        
  filecolor=blue,      
  urlcolor=blue           
}

\begin{document}
\title{Influence of the photonuclear effect on electron-neutrino-induced electromagnetic cascades under the Landau-Pomeranchuk-Migdal regime in standard rock}

\author{Mathieu Tartare}\email[corresponding author : ]{mathieu.tartare@lpsc.in2p3.fr}   \affiliation{Laboratoire de Physique Subatomique et de Cosmologie (LPSC), Universit\'{e}
Joseph Fourier, INPG, CNRS-IN2P3, Grenoble, France}
\author{Didier Lebrun}  \affiliation{Laboratoire de Physique Subatomique et de Cosmologie (LPSC), Universit\'{e}
Joseph Fourier, INPG, CNRS-IN2P3, Grenoble, France}
\author{Fran\c cois Montanet} \affiliation{Laboratoire de Physique Subatomique et de Cosmologie (LPSC), Universit\'{e}
Joseph Fourier, INPG, CNRS-IN2P3, Grenoble, France}


\begin{abstract}
The observation of earth skimming neutrinos has been proposed as a rather sensitive method to detect ultra-high energy (UHE) cosmic neutrinos. Energetic cosmic neutrinos can interact inside the rock and produce leptons via a charged current interaction. In the case of an incoming $\nu_e$ undergoing a charged current interaction, the produced UHE electron will induce an underground electromagnetic shower. At high energy (above 7.7 TeV in standard rock), such showers are subject to LPM (Landau, Pomeranchuk and Migdal) suppression of the radiative processes cross sections (bremsstrahlung and pair production). The consequence of this suppression is that showers are elongated. This effect will increase the detection probability of such events allowing deeper showers to emerge with detectable energies. On the other hand, the photonuclear processes which are usually neglected in electromagnetic showers with respect to radiative processes, turn out to become dominant in the LPM regime and will reduce the shower length. In this work, we have performed a complete Monte Carlo study of an underground shower induced by UHE electrons by taking into account both the LPM suppression and the photonuclear interaction. We will discuss the effects of both of these processes on the shower length and on the detectability of such events by ground arrays or fluorescence telescopes. We show that limits on neutrino fluxes that were obtained using simulations that were obviously neglecting photonuclear processes are overoptimistic and should be corrected.

\end{abstract}
\maketitle

\section{Introduction}
The search for ultra-high energy (UHE) (E$>10^{17}$~eV) neutrinos requires a huge volume of target materials because of their low interaction probability and low fluxes. 
The earth provides different target materials: some experiments are
searching for UHE $\nu$ interacting in sea water (for example ANTARES
\cite{antares}), others in ice (IceCube \cite{icecube,icecube40,icecube40e} and ANITA \cite{anita,anitae}) or in the atmosphere (HiRes \cite{hires} and the Pierre Auger Observatory \cite{nutau2,down}). 
One can also search for UHE $\nu_\tau$ interacting in the earth's crust
in trying to detect $\tau$ decay-induced showers \cite{tauastropart}.  
Indeed, up-going $\nu_\tau$ can interact near enough to the earth's
surface to produce a $\tau$ whose range is long enough to escape the
earth, decay and produce a shower in the atmosphere \cite{nufargion}. This shower can trigger a surface detector array such as the Pierre Auger Observatory Surface Detector~(SD)~\cite{SD1}. 


Another way of detecting up-going neutrinos is via the interaction of $\nu_e$ into the earth's crust. 
UHE $\nu_e$ will interact via charged current (CC) interaction with a nucleus of the medium and produce an electromagnetic and a hadronic shower. Since most of the incident neutrino energy is transferred to the electron ($\gtrapprox 80\%$ of $E_\nu$), the electromagnetic shower will be very energetic. 
The hadronic shower produced by the neutrino CC interaction will not contribute significantly to the detection probability because the energy transferred to it is negligible. In this paper we will discuss the development in rock of the $\nu_e$ induced electromagnetic showers.

An important effect plays a role in electromagnetic showers at high energy: the Landau, Pomeranchyk, and Migdal (LPM) effect, first predicted by Landau and Pomeranchuk \cite{landau} and derived in a modern quantum formalism by Migdal \cite{migdal,classicallpm}. 
This effect induces an increase of the radiation length and hence the shower will be more penetrating and will have a higher probability to emerge in the atmosphere with large enough energy to be detected by a surface detector. 

Another effect needs to be taken into account which limits the LPM
interaction length enhancement at high energy: the photonuclear
interaction. The photonuclear effect consists of the interaction of a
photon with a nucleus of the medium (via a direct or resolved process)
which will produce a hadronic sub-shower. This effect, whose cross
section increases logarithmically with energy of the photon, will increase the photon total cross section and therefore reduce its mean free path. On the other hand, the produced hadronic subshower will not propagate too far because of the small hadronic interaction length in rock. 

In this paper, we first start by presenting the LPM and the photonuclear cross sections and their energy dependence. We then discuss the consequences of these two effects on underground electromagnetic showers at UHE. 
We then illustrate this showing the results of a Monte Carlo study of the development of realistic showers in rock.

We will conclude this work by studying the shower length and the probability that the emerging shower tail exceeds a given detection threshold. In this way we get a sense on their detectability by ground arrays or fluorescence detectors. We show that by neglecting photonuclear processes one overestimates the detection sensitivity by a large factor and that earlier neutrino limits that were inferred in this manner are probably overoptimistic.

\section{Electron and photon interactions at UHE}
Bethe and Heitler first calculated the cross sections of pair production and Bremsstrahlung. 
Some corrections were then added to these cross sections to take into account screening effects. 




At UHE, the bremsstrahlung process takes place within a relatively
large distance which is known as the photon formation length $l_f$ and
is given by $l_f = \hbar/q_{||}$ where $q_{||}$ is the longitudinal
momentum transferred to the target nucleus by an electron with energy
$E$ and mass $m$ emitting a photon with energy $k$:
$q_{||}=km^2c^3/2E(E-k)$, where $c$ is the speed of light. $l_f$ rises with electron energy. 
When $l_f$ becomes larger than the interatomic spacing, then the amplitudes of multiple interactions with multiple atoms add coherently and destructive interferences occur and suppress the bremsstrahlung amplitude (note that this was not taken into account in the original Bethe-Heitler theory). 
A similar reasoning applies to the pair production process \cite{klein2}.
The bremsstrahlung suppression is significant for the emitted photon with energy $k$ from an electron with energy $E$ when
\[
k < \frac{E(E-k)}{E_{\mathrm{LPM}}},
\]
with $E_{\mathrm{LPM}}$ given by
\[
E_{\mathrm{LPM}}= \frac{m^4X_0}{E_s^2},
\]
with $E_s = mc^2\sqrt{4\pi/\alpha} = 21.2$ MeV, where $\alpha$ is the
fine-structure constant. In standard rock (constant density $\rho = 2.65$~g.cm$^{-3}$, $X_0= 27.6$~g.cm$^{-1} \equiv 10.4$ cm), one gets $E_{\mathrm{LPM}} = 77$~TeV so that in our case where we consider $E \gg$~TeV, this effect cannot be neglected.
The modification of the electron interaction length under LPM regime can be approximated by
\[
X_e = X_0\sqrt{\frac{2E}{E_{\mathrm{LPM}}}},
\]
and for the photon by: 
\[
X_\gamma = X_0 \sqrt{\frac{k}{50E_{\mathrm{LPM}}}}.
\]
For example, an electron with an energy of 10$^{18}$~eV will have its
interaction length increased from 0.1~m to 22~m in standard rock
because of the LPM suppression: the LPM effect will increase the
shower length as mentioned in \cite{nueshower}.
A purely electromagnetic shower under the LPM regime will develop over a much larger distance.

Another process, playing a dominant role at high energy, is the photonuclear effect. Photons can interact with nuclei of the medium and produce a hadronic subshower. At these energies, a photon interacts either directly with a parton of the nucleus or may fluctuate into a quark antiquark state which will interact with the nucleus. 
Models based on different descriptions give fairly different cross sections for $\gamma p$ interactions. One can cite, for example, models such as in \cite{GVDM} assuming generalized vector meson dominance together with direct photon-quark interaction which takes into account resolved and direct photon interactions, model based on VDM (vector meson dominance) with a Gribov-Glauber approach \cite{kleinphotoprod,Glauber} or the Donnachie-Landshoff model (exchange of pomeron pairs) in the Regge theory framework \cite{regge}. 
A comparison between Gribov-Glauber and GVDM may be found in \cite{kleinphotoprod}.

In lower energy showers, this process is hidden behind the dominant radiative processes. In the LPM regime, this interaction becomes important as the cross section for hadronic interactions ($\gamma p $ interactions) becomes larger than the cross section for pair production. The crossover energy between these two processes is between $10^{19}$ and $10^{20}$~eV and is found to be more or less independent of the material. This may be explained by the fact that the increase in $\sigma_{\gamma p}/\sigma_{ee}$ for heavier nuclei is canceled out by the decrease of $E_{LPM}$ as $X_0$ drops, as underlined in \cite{klein2,nueshower}. This crossover energy is highly dependent on the model used to calculate the photonuclear cross section.

This cross section rise reduces the mean free path of photons in the electromagnetic shower and therefore reduces the total shower length. In addition, a hadronic and muonic component appears in the shower which will no longer be purely electromagnetic. Part of the energy of the electromagnetic shower leaks to this hadronic component and this will further decrease the shower length.

Our goal here is to study the shower length and energy under these two processes via a complete and realistic Monte Carlo simulation of underground electron-induced showers. 
One can mention that the need for detailed simulation of photonuclear interactions was already mentioned in \cite{kleinradio}.

In our study we used TIERRAS \cite{tierras} to simulate underground
showers. TIERRAS is an adaptation of AIRES (Air-shower Extended
Simulations) \cite{userguideaires} for material like rock or ice. This
simulation code provides a complete and realistic Monte Carlo
treatment of the interactions which takes place within the shower
development in dense matter. It implements Migdal's theory of the LPM
effect exactly and solves recursively the Migdal's cross section
parameters \cite{migdal}.

The definition of standard rock used in this study refers to rock with constant density $\rho = 2.65$~g.cm$^{-3}$ (density of continental earth crust),
effective $Z=11$ and $Z/A=0.5$ (i.e., isoscalar nuclei)~\cite{tierras}. The density of the earth crust used by the HiRes collaboration in \cite{hires} is $2.80$~g.cm$^{-3}$. We do not expect this small discrepancy to impact our conclusions because the cross over energy between PN and LPM cross sections is essentially independent of the density. 


The photonuclear interactions are processed with QGSJETII \cite{QGSJET} and the cross section is given by a parametrization and extrapolation of experimental data given in \cite{baldini}. One should note that QGSJETII does not take into account the production of charmed mesons and associated prompt muons production.

\begin{figure} []
\centering

\includegraphics[width=0.5\textwidth]{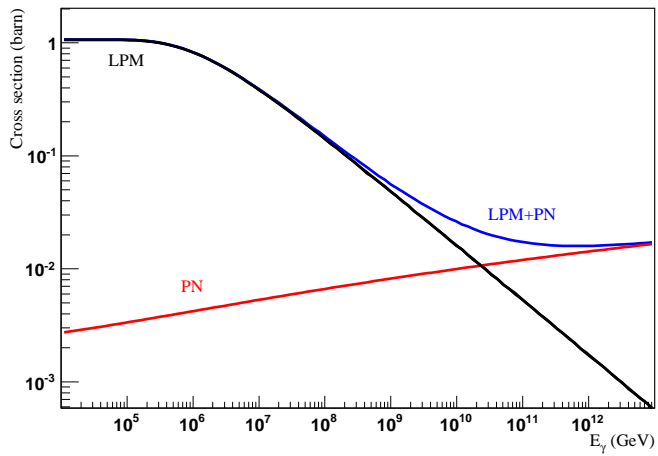}

\caption
{Pair production under LPM regime (LPM, black) and photonuclear
  interaction (PN, red) cross sections in standard rock used in TIERRAS~\cite{tierras}.}
\label{fig:xsection}
\end{figure}

Cross sections used in this study are shown in Fig.\ref{fig:xsection}\\

As shown in \cite{klein2}, another effect plays a role in the underground shower development: the electronuclear interaction. This effect is currently not taken into account by TIERRAS but it would be interesting to implement it in the future.

\section{Monte Carlo study}

Using TIERRAS, we simulated underground showers induced by electrons
with an energy between $10^{17}$~eV and $10^{20}$~eV (which is the
range where the transition between the LPM and PN cross sections is
expected). In these simulations, we propagate the showers first using
standard bremsstrahlung and pair production processes (later called
Bethe-Heitler, ''BH case''), then accounting for the LPM effect (''LPM
only'' case) and finally enabling the photonuclear effect (''LPM+PN
case'').

The hadronic component produced by photonuclear interactions is also simulated even though we do not expect it to propagate over large distances: we thus propagate and follow electrons, photons, muons and hadrons.

As in the case of air-shower simulations, for UHE primaries, the number of particles that are produced is large enough to make it impossible to propagate  all the secondaries. The simulations are made possible thanks to a sampling algorithm: the Hillas thinning algorithm~\cite{hillas}. One of the drawback of this algorithm is to introduce artificial fluctuations in the distribution of different observables such as the longitudinal profile, the lateral distribution, or the energy of the shower. In order to get smoother distributions we used a low thinning factor of $10^{-6}$ (given by $E_{\mathrm{th}}/E_{\mathrm{prim}}$, where $E_{\mathrm{th}}$ is the energy below the thinning algorithm is applied and $E_{\mathrm{prim}}$ the energy of the primary). In order to control the fluctuations induced by sampling, we also set the statistical weighting factor $W_f$ to 0.2~\cite{userguideaires}. For more information about the impact of the thinning algorithm on shower simulations, see \cite{userguideaires}. 
\begin{figure} [tb]
\centering
\subfigure{
\includegraphics[width=0.5\textwidth]{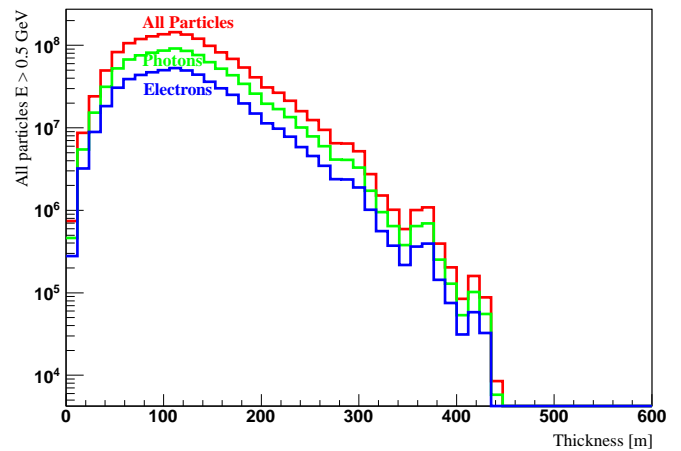}}
\subfigure{
\includegraphics[width=0.5\textwidth]{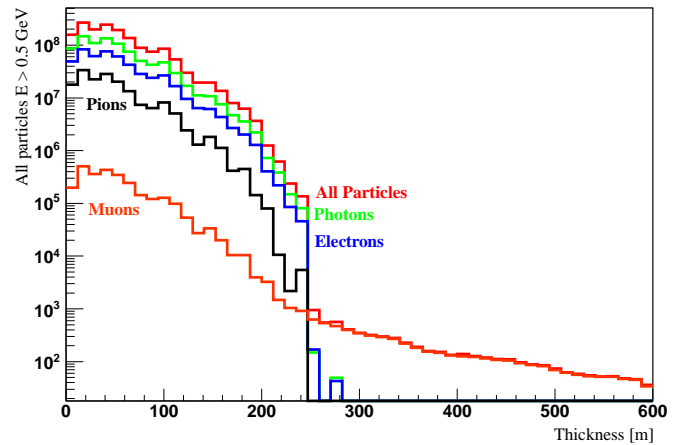}}
\caption{Longitudinal Profile for shower induced by a 100~EeV electron
  with ``LPM only''in the top and with ``LPM and photonuclear''
  processes in the bottom. The all particle profile is shown in red. The different components, namely electrons, photons, pions and muons are respectively displayed in blue, green, black and yellow.}
\label{fig:LongProfile}
\end{figure}

\begin{figure} 
\centering

\includegraphics[width=0.5\textwidth]{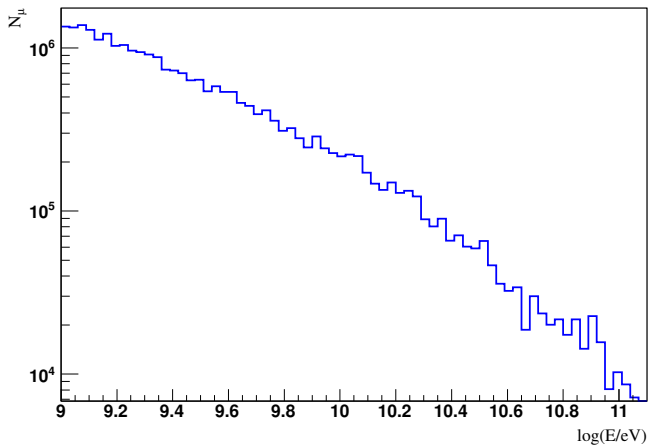}
\caption{Energy spectrum of muons at the mean maximum development of the underground shower}
\label{fig:muspect}
\end{figure}

On Fig.\ref{fig:LongProfile} one can see the longitudinal profile for
$10^{20}$~eV electron showers, averaged over 100 showers, for each
type of particle, and for the following cases:
 with LPM suppression only (''LPM only'') and with LPM and photonuclear effects (''LPM+PN''). 
One can see that the "LPM only" showers will develop along much larger distances than the "LPM+PN" showers. One can also see the apparition of hadrons and muons in the shower in the "LPM+PN" case highlights the fact that a part of the total shower energy is transmitted to hadronic subshowers and that a muonic component builds up and propagates much further 	away than the rest of the shower and carries a non negligible part of the total shower energy. This is of course of interest for an experiment sensitive to muons such as the Pierre Auger Observatory. Figure\ref{fig:muspect} shows the energy spectrum of muons above $1$~GeV at the maximum of development of the shower. Taking into account charmed meson production producing prompt muons would harden this spectrum.

\begin{figure}[tb]
\centering
\subfigure{
\includegraphics[width=0.5\textwidth]{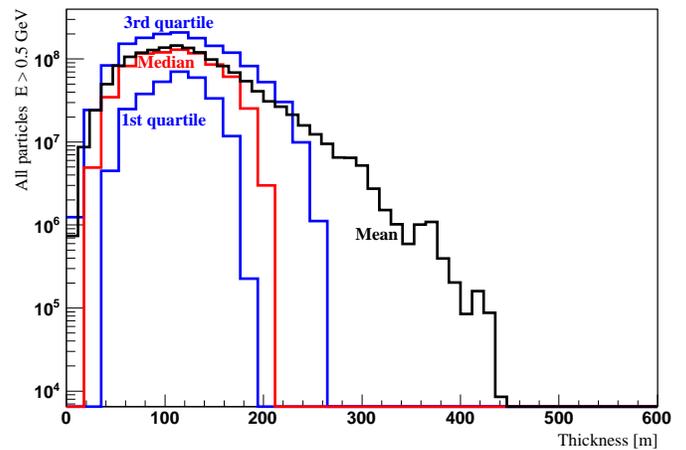}}
\subfigure{
\includegraphics[width=0.5\textwidth]{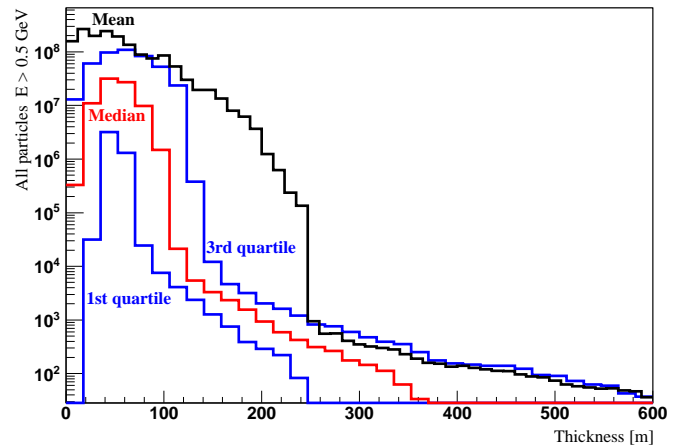}}
\caption{ Mean longitudinal profile (black), median (red), first and third quartile (blue) for shower induced by a 100~EeV electron
  with ``LPM only''in the top and with ``LPM and photonuclear''
  processes in the bottom.}
\label{fig:centile}
\end{figure}

If one would look at the longitudinal profiles of individual showers one would see important fluctuations from shower to shower. This is a characteristic behavior of electromagnetic showers in the LPM regime, as underlined in \cite{nueshower} and indeed confirmed by our simulation. As we can see on Fig.\ref{fig:centile}, there is an important difference between the mean longitudinal profile and its median. This is due to the fact that the mean is dominated by a few extreme values coming from a few showers. If one looks at the first and third quartile, we can see that the interquartile range is large: $\sim 1$~decade at the shower maximum development for the LPM case and up to $\sim 2$~decades for the LPM+PN case where the fluctuations are more important close to the maximum. This shows that one should be careful with the interpretations of the mean longitudinal profiles. Afterward, in order to avoid bias by averaging on all the showers, our calculations of the probability of an underground shower emerging in the atmosphere are made on a shower to shower basis.

\begin{figure} []
\centering

\includegraphics[width=0.5\textwidth]{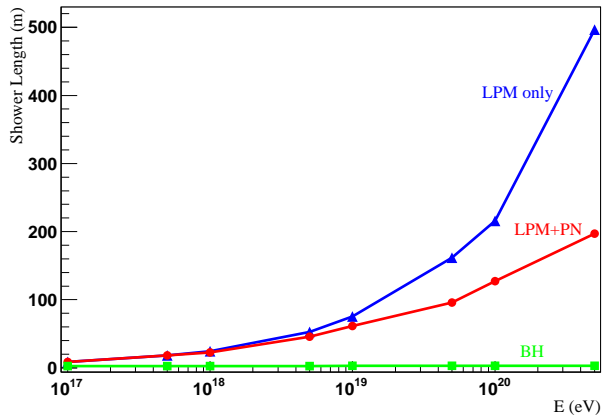}
\caption
{Shower length versus primary electron energy. Including
  ''Bethe-Heitler'' (squares), ''LPM only'' (triangles), ''LPM+PN'' (dots).}
\label{fig:length}
\end{figure}

In order to estimate the shower length we need to define an estimator of the latter.
We adopted the definition of the ``shower length'' as given by Alvarez-Mu\~{n}iz and Zas in \cite{alvarez} where it is defined as the length over which the number of particles in the shower is larger than an arbitrary fraction of the maximum number of particles in the shower. We choose here to set this fraction to $1/10$. Results are shown in Fig.\ref{fig:length}, where we plot the shower length as a function of the energy of the primary electron.

One can notice that photonuclear processes make no difference in the shower length below $10^{18}$~eV because the photonuclear cross section is dominated by radiative processes at these energies as can be seen from the cross sections (Fig.\ref{fig:xsection}). The shower length increases strongly with energy when the LPM effect is applied without photonuclear effect. This increase is severely reduced when both LPM and photonuclear effects are taken into account. 

These results are in good agreement with \cite{klein2} in which the same behavior is predicted from a phenomenological approach.


In a second study, we then consider the probability that the EM energy of the emerging shower tail exceeds a given detection threshold. A reasonable estimate of what would be the energy threshold for an array of surface detectors such as the Pierre Auger SD for detecting quasi-horizontal grazing showers emerging inside the array boundaries should be close to the $10^{17}$~eV. We thus fix the energy detection threshold to this value, although more realistic results would require a complete detector simulation. 

We used the same set of simulated showers as above (i.e. showers from primary electrons with energy between $10^{17}$ and $10^{20}$~eV). 

\begin{figure} []
\centering

\includegraphics[width=0.5\textwidth]{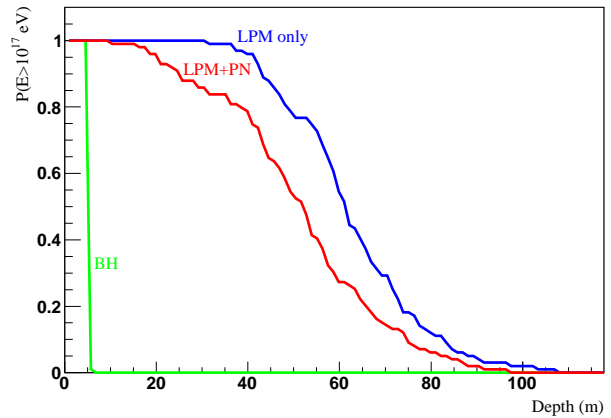}
\includegraphics[width=0.5\textwidth]{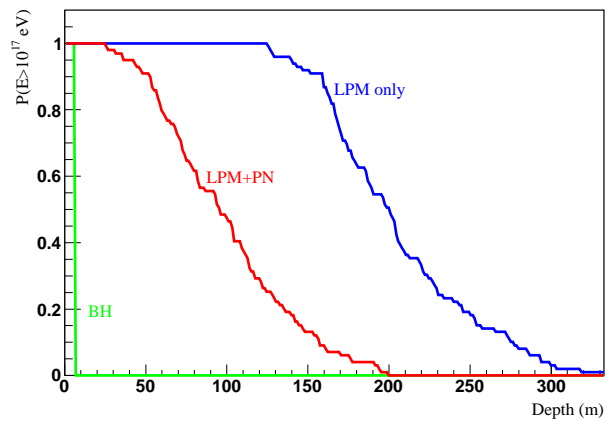}

\caption{Probability of an underground shower with primary energy of
  $10^{18}$~eV (in the top) and $10^{20}$~eV (in the bottom) to emerge
  with an energy above $10^{17}$~eV after crossing a given amount of
  rock (in meters on this graph) for the three cases: ``BH'', ``LPM
  only'' and ``LPM+PN''.} 
\label{fig:ProbaEmerge}
\end{figure}

On Figure \ref{fig:ProbaEmerge} we represent the probability that an
underground shower with primary energy of $10^{18}$~eV and $10^{20}$~eV emerges with a remaining energy above $10^{17}$~eV after crossing a given amount of rock.

One can see that the probability for a shower emerging with an energy
above $10^{17}$~eV is higher in the ``LPM only'' case than in the two
other cases. Photonuclear processes increase the energy loss rate in a
non negligible way. If we compare the ``LPM only'' and ``LPM+PN'' case
one can notice that at $10^{20}$~eV the total probability is up to 2 times
higher in the LPM case while at $10^{18}$~eV this total probability is only
$\sim$ 1.2 times higher. At $10^{18}$~eV the overestimation of the
emerging probability is not really important, but above $10^{19}$~eV,
PN interactions start to make a difference between the ``LPM only''
and ``LPM+PN'' case. 

In \cite{hires}, the High Resolution Fly's Eye collaboration published a limit for up-going $\nu_e$ inducing LPM showers in crust. In their article, the authors define $\epsilon_t$ as the transmission probability, which is the probability for a shower to emerge into the atmosphere with more than $10^7$ particles (above a low energy threshold). This probability can easily be extracted along with the shower length in our simulations. The energy dependence we obtain for $\epsilon_t$ shows the same trends as the curves corresponding to the mean shower length. 
This probability is equivalent to the ratio between the mean shower
length and the total thickness of rock considered (by defining the
shower length as the length where the shower has more than $10^7$
particles). Here we are considering a total rock thickness of 600 m
and since we do not know the energy threshold used by the HiRes
collaboration, we choose to fix this threshold at $500$~MeV regardless
of the type of the particle, assuming
that particles with lower energy will not contribute significantly to
the detection.

\begin{figure} []
\centering

\includegraphics[width=0.5\textwidth]{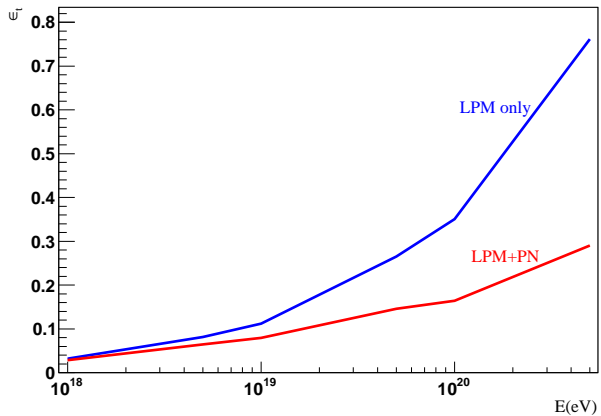}

\caption{Probability of an underground shower to emerge with a number of particle above $10^7$ versus primary energy, normalized to 600 m of rock. ``LPM only'' in blue and ``LPM+PN'' in red.}
\label{epsilon}
\end{figure}



By comparing the values of $\epsilon_t$ in the ``LPM only''  case and in
the ``LPM+PN'' case, one can say that neglecting PN interaction at
$10^{18}$~eV only leads to an overestimation of $\epsilon_t$ of $\sim 10\%$, at $10^{19}$~eV this
overestimation grows to $\sim 40\%$ and at $10^{20}$~eV to $\sim 110\%.$

One concludes that neglecting the photonuclear
effect causes an overestimation of the emerging shower energy and of
the emerging number of particles above $10^{19}$~eV. The detection
probability is related to $\epsilon_t$, which is itself related to the
total energy and number of particles emerging into the
atmosphere. Overestimating the emerging energy as well as the shower
length will lead to an artificial enhancement of $\epsilon_t$ above
$10^{19}$~eV and therefore of the experimental sensitivity to this type of event.





\section{Conclusion}

Accounting for photonuclear interactions into our simulations, we come
to the conclusion that the LPM effect still enhances the shower length
in rock but this enhancement is limited in a non negligible way. The
underground shower energy loss is also impacted by the photonuclear
interactions, as we can see in the decrease of the probability to
emerge above a given energy. In counterpart a non negligible muonic
component will propagate over large distances under rock carrying a
part of the shower energy. It appears that one cannot
neglect the photonuclear effect in simulations of underground showers
where its cross section dominates the radiative processes cross
sections above $10^{19}$~eV. 

The development length of $\nu_e$ induced electromagnetic showers, the
emerging shower energy and therefore the exposure to up-going $\nu_e$
event have been overestimated in \cite{hires} (and more particularly
above $10^{19}$~eV. 
The corresponding $\nu_e$ flux limit should thus be reevaluated by the authors by taking photonuclear processes into account. 



Another important fact may impact further the development of LPM
showers. As outlined by some authors \cite{klein2,lpmquenching}, the
radiated bremsstrahlung photon may interact via the photonuclear effect
before being actually fully formed and this could further limit the
the LPM interaction length. We have not yet implemented this effect into
our simulation and it is so far not taken into account in TIERRAS,
AIRES and CORSIKA~\cite{corsika} codes. Taking into account this effect would further reduce the shower length and hence increase the discrepancy between LPM case and a treatment including all processes.


Further work is hence needed to account in full detail for emerging showers and evaluate the sensitivity of UHE cosmic ray observatories to the flux of $\nu_e$ searching for these intriguing events.

\section{Acknowledgments}
We thanks Jaime Alvarez-Mu\~{n}iz and Sergio Navas for their useful
comments and Matias Tueros who developed TIERRAS \cite{tierras}. 


\end{document}